\documentclass[aps,prl,twocolumn,showpacs,superscriptaddress,floatfix,citeautoscript,floatfix]{revtex4-1}

\usepackage{dcolumn}
\usepackage{bm}
\usepackage{amsmath,amssymb}
\usepackage{times}
\usepackage{float}
\usepackage{graphicx, epstopdf}
\usepackage{color}
\usepackage{setspace}
\usepackage{natbib}

\begin{document}

\title{Accurate, rapid identification of dislocation lines in coherent 
diffractive imaging via a min-max optimization formulation}

\author{A. Ulvestad}
\affiliation{Materials Science Division, Argonne National Laboratory, Lemont, 
IL 60439, USA}
\author{M. Menickelly}
\affiliation{Mathematics and Computer Science Division, Argonne National 
Laboratory, Lemont, IL 60439, USA}
\author{S. M. Wild}
\affiliation{Mathematics and Computer Science Division, Argonne National 
Laboratory, Lemont, IL 60439, USA}

\begin{abstract}
Defects such as dislocations impact materials properties and their
response during external stimuli. Defect engineering has 
emerged as a possible route to improving the performance of materials over a wide 
range of applications, including batteries, solar cells, and 
semiconductors. Imaging these defects in their native operating 
conditions to establish the structure-function relationship and, ultimately, to 
improve performance has remained a 
considerable challenge for both electron-based and x-ray-based imaging techniques. 
However, 
the advent of Bragg coherent x-ray diffractive imaging (BCDI) has made possible 
the 3D imaging of multiple 
dislocations in nanoparticles ranging in size from 100 nm to1000 nm. While the imaging 
process succeeds in many cases, nuances in identifying the dislocations has left 
manual identification as the preferred method. Derivative-based methods are also 
used, but they can be inaccurate and are 
computationally inefficient. Here we demonstrate a derivative-free method that 
is both more accurate and more computationally efficient than either derivative- or 
human-based methods for identifying 3D 
dislocation lines in nanocrystal images produced by BCDI. We formulate the 
problem as a min-max 
optimization problem and show exceptional accuracy for experimental images. We 
demonstrate a 260x speedup for a typical experimental dataset with higher 
accuracy over current methods. We discuss the possibility of using this 
algorithm as part of a sparsity-based phase retrieval process. We also provide 
the MATLAB code for use by other researchers. 
\end{abstract}

\maketitle 
In materials, crystallographic imperfections such as dislocations often 
dictate performance and properties. For example, dislocation cores can act as 
fast diffusion sites \cite{Heuser2014, PurjaPun2009, Xiong2014, Li2014}, 
mitigate strain and plasticity during structural phase transformations 
\cite{Ulvestad2017,yaunano,Gaucherin2009}, and govern crystal growth 
\cite{Frank1949, Clark2015a,Walker2004}. Increasingly, Bragg coherent x-ray 
diffractive imaging (BCDI) is being utilized at synchrotron and x-ray free 
electron laser \cite{Clark2013,Ulvestad:2017aa} sources to address 
this challenge of understanding and optimizing materials properties via tuning of 
lattice distortions by nondestructively imaging the 3D lattice distortion field 
under \textit{in situ} and \textit{operando} conditions \cite{Robinson2009, 
Cha2013, Watari2011b, Miao2015, Hofmann2017b, Newton2010, Cherukara2016}. The 
typical resolution of the technique is roughly $10^{-4}$ in strain sensitivity 
and 
10--20 nm spatial resolution at 5--10 minute temporal resolution. Recent 
studies have revealed the 3D dislocation line dynamics in an individual 
nanocrystal during reactive processes such as hydrogen uptake 
\cite{Ulvestad2017, yaunano}, battery charging \cite{Ulvestad2015a}, crystal 
growth and dissolution \cite{Liu2017, Clark2015a}, and grain growth in 
polycrystalline materials \cite{Yau2017}. In all these studies, accurate 
identification of the dislocation line in the reconstructed image is essential 
to understanding the underlying physics and the dislocation impact. 
While these studies have shown great promise, the breadth of 3D BCDI dislocation 
dynamics measurements and techniques could expand substantially if accurate, 
robust, and rapid methods existed to determine the 3D dislocation line 
structure. For example, the datasets generated at diffraction-limited storage 
rings will likely be too large for existing derivative-based and 
human-in-the-loop-based methods 
\cite{Dietze2015}, and the ``dislocation basis" could potentially be used as a 
sparse basis to circumvent constraints in phase retrieval \cite{Tripathi2016}. 
Here we present such a method by reformulating the dislocation core 
identification problem as a min-max optimization problem that can be solved 
rapidly.

It is counterintuitive that an imaging experiment with 10--20 nm spatial 
resolution is sensitive to atomic-scale defects such as dislocations. To 
understand how this is possible, consider the relationship between the continuum 
representation of the crystal, $\rho(\mathbf{r})$, and the diffraction 
intensity, $I(\mathbf{q})$ in the far field under a perfectly coherent 
illumination and in the kinematical scattering approximation 
\cite{Vartanyants2001,Robinson2009,Als-nielsen2011}:
\begin{eqnarray}
I(\mathbf{q})\approx \left|\mathcal{F}\left(\rho(\mathbf{r})e^{i 
\mathbf{Q}\cdot \mathbf{u}(\mathbf{r})}\right)\right|^2.
\label{mod}
\end{eqnarray}
Here, $\mathbf{r}$ and $\mathbf{q}$ are the real and reciprocal space 
coordinates, respectively, $\mathcal{F}$ is the Fourier transform that 
provides the map between these two spaces, 
$\mathbf{Q}$ is the measured reciprocal lattice vector (e.g., the 111 Bragg 
peak for a face-centered cubic crystal lattice), and $\mathbf{u}(\mathbf{r})$ 
is the vector displacement field that is a continuum description of how the 
atoms are displaced from their equilibrium positions. If we consider a cubic 
crystal with a screw dislocation along the $z$ direction, then the displacement 
field is given by $u_x=u_y=0$ and 
\begin{eqnarray}
u_z = \frac{b}{2\pi}\theta,
\end{eqnarray}
where $b$ is the Burgers vector and $\theta$ measures the angle around the 
dislocation core \cite{Hull2011}. If the Bragg peak (reciprocal lattice vector) 
measured is the $001$ peak, then 
$\mathbf{Q}\cdot \mathbf{u}(\mathbf{r})=|Q_{001}| \frac{b}{2\pi}\theta = 
\frac{2\pi}{a}\frac{b}{2\pi}\theta$. The Burgers vector in this case is equal 
to the lattice spacing in the $z$ direction, and so $\mathbf{Q}\cdot 
\mathbf{u}(\mathbf{r})= \theta$. Thus, the signature of a dislocation in the 
complex image is a point around which the phase varies from $0$ to $2\pi$, or 
$-\pi$ to $\pi$ depending on the chosen convention. This is the maximum 
``signal" 
possible in the image phase and is a direct consequence of 
dislocations introducing large displacement fields. Note that this argument can 
be extended to other Bragg peaks that are not parallel to the displacement 
field induced by the dislocation.

\begin{figure*}
\includegraphics[width=\linewidth]{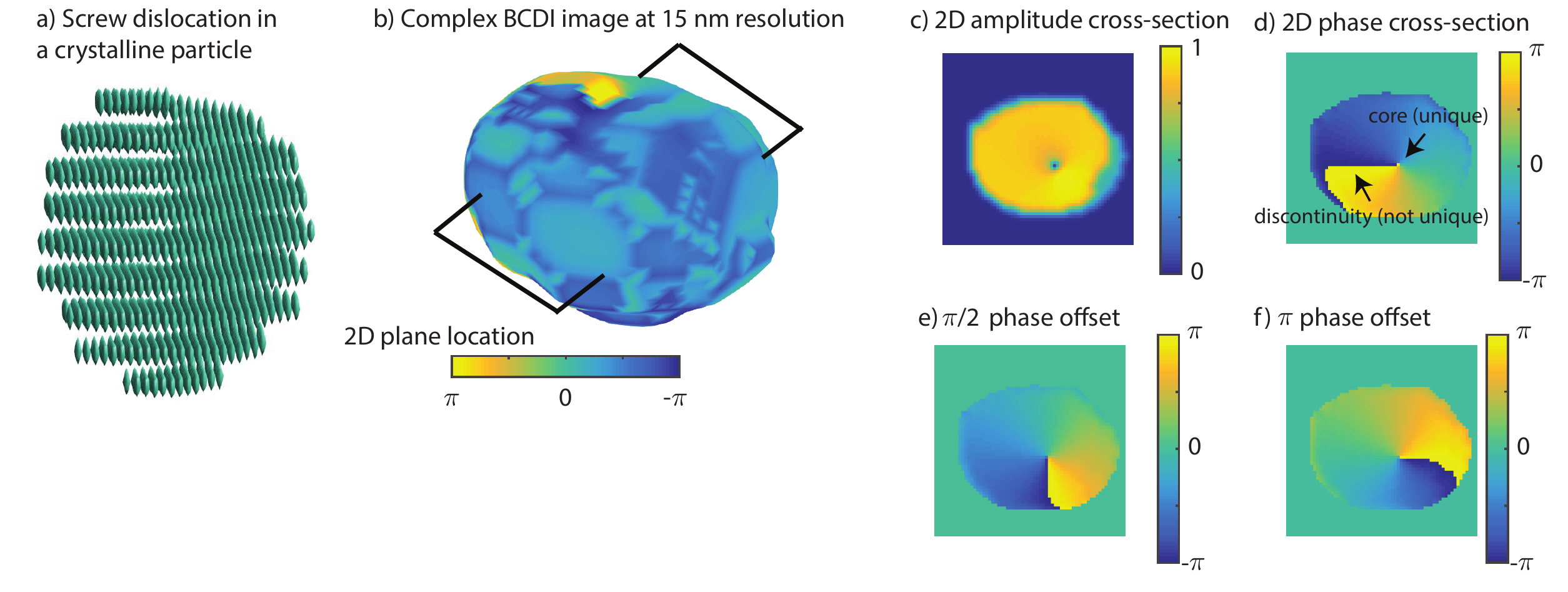}
\caption{
\label{fig:blur} 
Imaging an atomic-scale defect without atomic resolution. (a) Atoms of a cubic 
crystal are displaced according to the screw dislocation theoretical 
displacement field. (b) The $001$ Bragg peak is used after Gaussian blurring to 
simulate finite resolution and, assuming perfect phase retrieval, to reconstruct a 
BCDI image of the crystal. The particle shape is represented as an isosurface of 
the amplitude, which is proportional to the Bragg electron density, while the 
color projected onto the shape shows the atomic displacement field. (c) The 2D 
amplitude cross-section corresponding to the location shown in (b). The 
dislocation core signature can be seen as a low-amplitude region. (d) The 2D 
phase cross-section; the dislocation manifests itself as a phase vortex, or a 
region where the phase varies from $-\pi$ to $\pi$. Note that the 
dislocation core is unique but the phase discontinuity is not. To see why, 
consider (e) and (f), which have two different global (constant) phase offsets 
applied; (e) shows a global phase offset of $e^{i\pi/2}$ while (f) shows a 
global phase offset of $e^{i\pi}$. Both images produce the same diffraction 
pattern according to Eq.~\ref{mod}. Thus, the reconstructed image is not 
sensitive to the absolute phase.  Different global phase offsets move the phase 
discontinuity around the dislocation core. This fact is used in the 
derivative-based dislocation identification method.
}
\end{figure*}

\begin{figure*}
\includegraphics[scale=0.8]{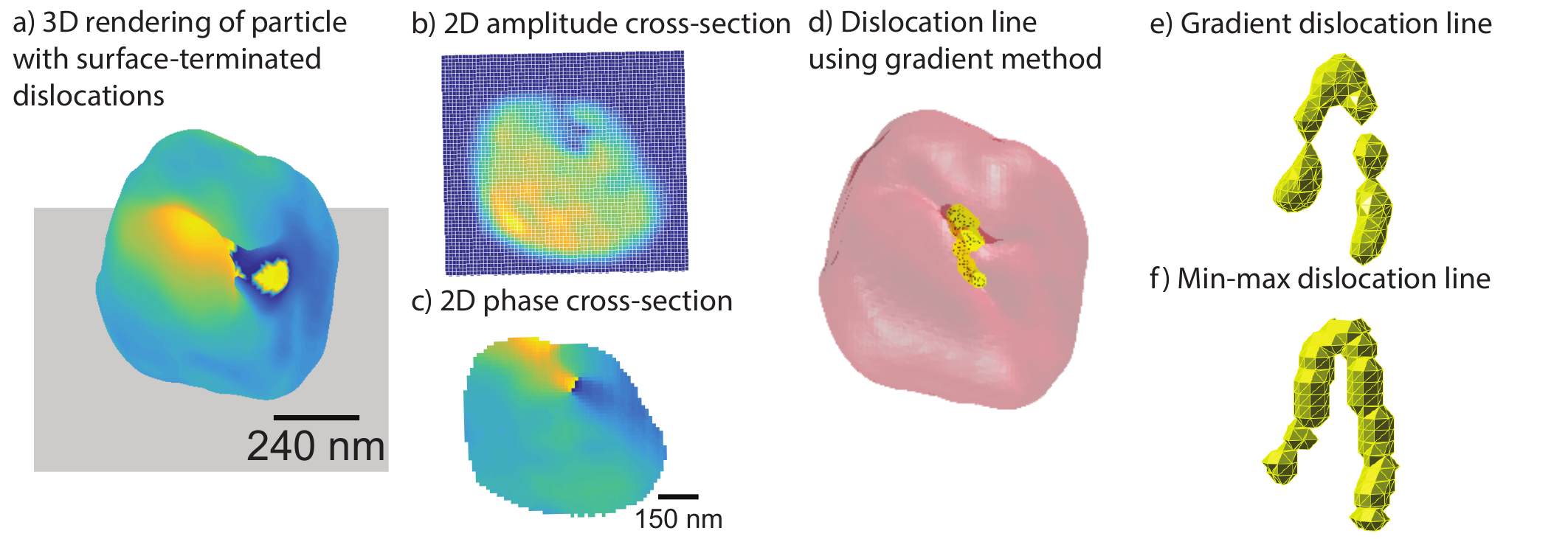}
\caption{
\label{fig:comp} 
Identifying the dislocation line in images reconstructed from experimental data 
in which a nanocrystal underwent electrochemical-induced dissolution. (a) The 
particle shape represented as an isosurface of the amplitude, which is 
proportional to the Bragg electron density, while the color projected onto the 
shape shows the atomic displacement field. The same color scale from $-\pi$ to 
$\pi$ is used as in Fig.~\ref{fig:blur}. (b) The 2D amplitude cross-section 
corresponding to the central vertical slice. The dislocation core signature can 
somewhat be seen as a low amplitude region. However, there are two dislocation 
cores and potentially three low amplitude regions so the identification is 
ambiguous. (c) The 2D phase cross-section. The two dislocation cores manifest 
themselves as phase vortices. (d) The 3D dislocation line (yellow) along with 
the particle shape (semi-transparent red) using the derivative-based method. (e) A 
close-up view of the dislocation line identified by using the derivative-based 
method. (f) A close-up view of the line identified using the min-max algorithm. 
The min-max algorithm is more accurate in identifying the line over all space and more computationally efficient. 
}
\end{figure*}

To demonstrate a BCDI study of a dislocation, we show in Fig.~\ref{fig:blur} 
the atomic and BCDI description of 
a screw dislocation in a cubic crystal. A 
periodic array of atoms is created to fill a particular particle shape, and the 
screw dislocation displacement field is applied (Fig.~\ref{fig:blur}a). The 
$001$ Bragg peak, blurred by a Gaussian function to simulate a finite 
resolution, is used to reconstruct the particle image at 15 nm resolution 
assuming perfect phase retrieval 
\cite{Marchesini2007a,Chapman2006a,Marchesini2003,Tripathi15}. The reconstructed 
BCDI image is complex. The amplitude corresponds to the Bragg, or diffracting, electron density 
\cite{Ulvestad2015b}, while the phase corresponds to the displacement of the 
atoms from their equilibrium positions. The amplitude is used to draw the 
isosurface while the phase is projected onto the isosurface as the colormap 
(Fig.~\ref{fig:blur}b). While the atoms cannot be identified in 
Fig.~\ref{fig:blur}b because of the finite resolution, there is a key signature 
in the phase in the form of a phase vortex, or a region where the phase varies 
from $-\pi$ to $0$ to $\pi$. The 2D cross-sections through the center of the 3D 
image are shown in Fig.~\ref{fig:blur}c-f. There is a signature of the 
dislocation core in the amplitude cross-section (Fig.~\ref{fig:blur}c), but 
this is often difficult to distinguish from the spatial amplitude variation in real 
experimental data (see, e.g., Fig.~\ref{fig:comp}b). The dislocation signature 
in the phase is much stronger 
(Fig.~\ref{fig:blur}d). Again, the reason is that dislocations are large 
displacements of the atoms from their equilibrium values. However, care must be 
taken in interpreting the displacement field image. In this case the 
dislocation core is 
unique. However, if periodicity is not appropriately accounted for, then 
the spatial location of the displacement field discontinuity, defined 
as the jump from $-\pi$ to $\pi$, is not unique, and so this discontinuity cannot 
be used directly for dislocation identification. To understand why, consider 
applying a global (constant) phase offset of $e^{i\pi/2}$ to the reconstructed 
image (Fig.~\ref{fig:blur}e). The diffraction pattern does not change because 
the factor is a constant and $|e^{i\phi_0}|=1$ for all $\phi_0$ (see 
Eq.~\ref{mod}). However, the phase discontinuity line shifts around the 
dislocation core. An additional example is shown in Fig.~\ref{fig:blur}f for a 
different phase offset. Thus, the phase discontinuity's spatial location is not 
unique, but the dislocation core's spatial location is unique. All phase offsets 
shift the discontinuity around the dislocation core, which is exploited in the 
standard derivative-based method for dislocation core identification described 
next. 

In previous work, dislocation cores were identified by eye or with a 
derivative-based method that considered a range of phase offsets as shown in 
Figs.~\ref{fig:blur}d-f. The derivative-based method was previously detailed in 
\cite{Clark2015a,Ulvestad2017}; only the key steps are repeated here. A 
derivative of the displacement field is taken in all three orthogonal image 
directions, thresholding is applied to determine what constitutes a large 
derivative value, and the locations of these large derivatives are 
stored. The process is 
repeated for 360 different global phase offsets in $1^{\circ}$ steps. 
The intersection (across these 360 offsets) of the locations with large 
derivatives is used to identify the 
dislocation core. This process is computationally and memory intensive (three 
derivatives for each voxel  for all 360 phase offsets need to be computed and 
stored) and requires tuning multiple thresholds. We now define the new algorithm 
and show how it is both more efficient and more accurate.

\begin{figure*}
\includegraphics[scale=.85]{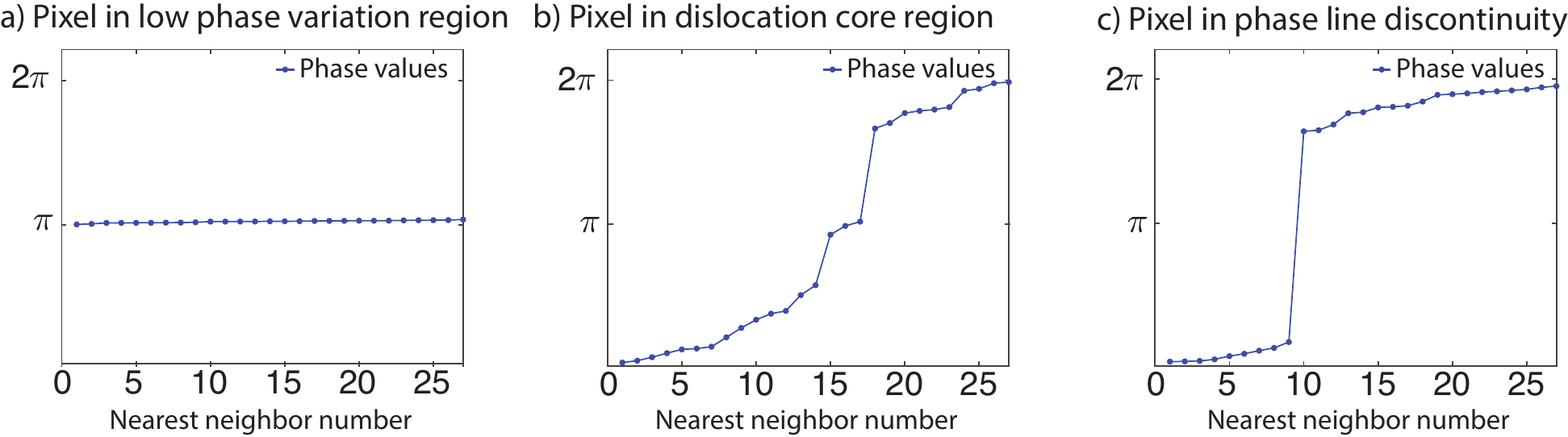}
\caption{
\label{fig:new} 
Sample phase values from a reconstruction of experimental data (remapped onto a 
$[0,2\pi]$ scale) for the 25 nearest neighbors of three pixels from 
distinctly different regions. (a) Sorted values in the neighborhood of a 
pixel showing minimal phase variation. (b) Sorted values in the 
neighborhood of a pixel in the dislocation core region. (c) Sorted values in 
the neighborhood of a pixel in the phase discontinuity region. In our algorithm, 
the value of $\Psi_{\alpha}$ in (a), (b), and (c) is roughly $0$, 
$4.2$, and $1.6$, respectively. This suggests a cutoff threshold of 4 
for identifying dislocation cores.}
\end{figure*}

We denote by 
$\hat{x}$ the integer-valued spatial coordinates of a pixel for which we 
have obtained
the complex-valued signal from phase retrieval, where we denote by 
$\rho(\hat{x})$ and $\theta(\hat{x})$ 
the amplitude and phase values at that pixel, respectively.
We further assume that, in preprocessing, a binary mask has been applied to 
$\theta$ so that a set of pixels $\hat{x}$ 
with low amplitude (i.e., $|\rho(\hat{x})|<\tau$ for some specified 
threshold $\tau>0$) are not considered.
We denote the set of all masked pixels $\hat{x}$ by $M$ and its
set complement by $\bar{M}$.
Such a mask has, for 
example, been applied to Figs.~\ref{fig:comp}b-c to mask the pixels that 
have near-zero amplitude. 
For arbitrary $\hat{x}$ and $\alpha\geq 0$, denote the 
\emph{$\alpha$-neighborhood of $\hat{x}$} by
$\mathcal{B}_{\alpha}(\hat{x}) = 
\{\hat{y}: \|\hat{x}-\hat{y}\|_{\infty}\leq\alpha\}$.
Consider the $\alpha$-parameterized 
function of phase offset $\phi_0$ and spatial coordinates $\hat{x}$
given by 
$$\begin{array}{rl}
   f_{\alpha}(\hat{x},\phi_0) \triangleq &  
\displaystyle\max_{\hat{y}\in\mathcal{B}_{\alpha}(\hat{x})\cap\bar{M}} 
\left[(\theta(\hat{y}) + \phi_0\right)\equiv 2\pi]\\
   - & \displaystyle\min_{\hat{y}\in\mathcal{B}_{\alpha}(\hat{x})\cap\bar{M}} 
\left[(\theta(\hat{y}) + \phi_0\right)\equiv 2\pi],
  \end{array}
$$
where we have used $\equiv$ to denote modulo (e.g., $[2\pi + 1 \equiv 2 
\pi]=1$). 
We adopt the convention that if $\mathcal{B}_{\alpha}(\hat{x})\cap\bar{M}=\emptyset$
(i.e., $\mathcal{B}_{\alpha}(\hat{x})\subset M$), then 
for any value of $\phi_0$, $f_{\alpha}(\hat{x},\phi_0)=0$.
Based on our previous discussion, large-magnitude values of the function
$$
  \Psi_{\alpha}(\hat{x}) \triangleq \displaystyle\min_{\phi_0\in[0,2\pi)} 
f_{\alpha}(\hat{x},\phi_0)
$$
signal that $\hat{x}$ is potentially at or near a dislocation core,
since $\Psi_{\alpha}(\hat{x})$ represents the least possible length (in radians)
of a radius 1 arc containing all the phases, modulo $2\pi$, of pixels in the 
neighborhood $\mathcal{B}_{\alpha}(\hat{x})$.
We are thus interested in solving the min-max problem
\begin{equation}
 \label{eq:minmax}
\max_{\hat{x}}  \Psi_{\alpha}(\hat{x}) = \max_{\hat{x}}
\displaystyle\min_{\phi_0\in[0,2\pi)} 
f_{\alpha}(\hat{x},\phi_0).
\end{equation}
We note that Eq.~\ref{eq:minmax}
is posed as a max-min problem, but we conform with the standard optimization 
terminology of ``min-max''
since any max-min problem has an equivalent min-max formulation obtained by 
negating the objectives. 
The interpretation of Eq.~\ref{eq:minmax} is that it seeks the largest values 
of phase differences under the best possible value of the phase offset $\phi_0$.

Because $\phi_0$ is a continuous parameter and $\hat{x}$ is a discrete pixel 
location, Eq.~\ref{eq:minmax} is a potentially challenging mixed-integer 
nonlinear 
robust optimization problem \cite{Belotti2013}. 
In our case, however, $\Psi_{\alpha}(\hat{x})$ is straightforward to evaluate.
We employ the finite list
$$\Theta(\hat{x}) \triangleq
\left\{[\theta(\hat{y})\equiv 2\pi]: 
\hat{y}\in\mathcal{B}_{\alpha}(\hat{x})\cap\bar{M}\right\}$$ 
of the phases associated with all unmasked neighbors of a pixel $\hat{x}$.
If $\Theta(\hat{x})$ is empty or composed of a single element, or if 
$\hat{x}\in M$,
then we follow the convention that $\Psi_{\alpha}(\hat{x})=0$.
Provided that $\hat{x}$ is not masked and that $\Theta(\hat{x})$ has at 
least two elements, we run the following procedure.
\begin{enumerate}
\item Sort the elements of $\Theta(\hat{x})$ in ascending order, 
$0 \leq \theta_1 < \dots< \theta_n < 2\pi$, 
where $n$ is the number of (distinct) elements in $\Theta(\hat{x})$.
\item Generate a length $n$ array $A$ defined by
$$
\begin{array}{rcll}
A_i & = & 2\pi-(\theta_{i+1}-\theta_i) & i=1,2,\dots,n-1 \\
A_n & = & \theta_n - \theta_1.			 &\\
\end{array}
$$
\item Return the least entry of $A$. 
\end{enumerate}

In practice, we typically set $\alpha=1$, as in Fig.~\ref{fig:new}.
We remark that in two dimensions, there are at most 9 phases in 
$\mathcal{B}_1(\hat{x})\cap\bar{M}$;
in three dimensions, there are at most 27 phases in 
$\mathcal{B}_1(\hat{x})\cap\bar{M}$.
Combining this information with the fact that any $\hat{x}\in M$ satisfies 
$\Psi_{\alpha}(\hat{x})=0$,
we can evaluate $\Psi_{\alpha}(\hat{x})$ at every unmasked pixel in 
a given 2D or 3D dataset in time that is linear in the 
number of unmasked pixels.
The min-max problem in Eq.~\ref{eq:minmax} is then trivially solved by taking 
the maximum of $\Psi_{\alpha}(\hat{x})$ over all such pixels $\hat{x}$, a task 
that can be performed in an online fashion (i.e., without explicitly storing 
the $\Psi_{\alpha}(\hat{x})$ values).

We can gain insight into the algorithm by considering the three cases shown in 
Fig.~\ref{fig:new}. Figure~\ref{fig:new}a shows the sorted experimental phase 
values in the neighborhood of a pixel in a region of minimal phase variation. 
The MATLAB range has been shifted from $[-\pi,\pi]$ to $[0,2\pi]$ by applying 
$+\pi$ to all values. These values were all close to zero in the reconstructed 
image, and thus they are all close to $\pi$ in the plot. Then, we compute $2\pi$ 
minus the pairwise differential ($A_i$ in step 2). This computation leads to values close to 
$2\pi$ for most of the pairs, since the differences are near zero. When 
computing the difference between the last point and the first point, the value 
is simply the difference, not $2\pi$ minus the difference ($A_n$ in step 2). The 
minimum of this set of values (step 3) is then close to zero, since point 27 and 
point 1 are close in value. This point is thus not likely to be near a 
dislocation core. 

Figure~\ref{fig:new}b shows the sorted experimental phase values in the 
neighborhood of a pixel in the dislocation core region. Again, we compute $2\pi$ 
minus the pairwise differential ($A_i$ in step 2) and the pairwise 
differential for point 27 and point 1 ($A_n$ in step 2). In this case, the 
minimum (step 3) in the $2\pi$ minus differential list comes from the 
difference between point 18 and point 17 (a difference of roughly $2$), and is 
thus roughly $4.2$. Pixel values above $\pi/2$ indicate potential proximity to 
a dislocation core, with this likelihood growing as the 
values approach the $2\pi$ upper bound on $\Psi_\alpha$.

Fig.~\ref{fig:new}c shows the sorted experimental phase values for a 
pixel outside of, but close to, the phase discontinuity region. In this case, 
the 
minimum in the $2\pi$ minus differential list comes from the difference between 
point 9 and point 10. The differential is approximately $3\pi/2$; since 
$2\pi-3\pi/2 = \pi/2$, the minimum value is approximately $1.62$. 

We compare our new method with the derivative-based method for identifying 
dislocations in images reconstructed from experimental data in 
Fig.~\ref{fig:comp}. Figure~\ref{fig:comp}a shows the experimental 
reconstruction of a silver nanoparticle after a dissolution step \cite{Liu2017}. 
The particle shape is represented as an isosurface of the amplitude, which is 
proportional to the Bragg electron density, while the color projected onto the 
shape shows the atomic displacement field. The large displacement field on the 
particle surface is due to the termination of two dislocation lines. A 
cross section of the amplitude (Fig.~\ref{fig:comp}b) and phase 
(Fig.~\ref{fig:comp}c) at the particle center show the signature of two 
dislocation lines. We used the derivative-based method to identify the 
dislocation line and show the line as a pair of yellow points superimposed onto 
the particle shape, which is shown as the semi-transparent red isosurface 
(Fig.~\ref{fig:comp}d). To better visualize the line, we show just the 
line in Fig.~\ref{fig:comp}e. The line clearly has a horseshoe-like 
shape, which is indicative of a dislocation transition between edge and screw 
character \cite{Hull2011}. The dislocation line shows some artifacts of the 
derivative-based method, namely, that some parts of the line appear to be larger 
or smaller than the voxels. Figure~\ref{fig:comp}f shows the line identified 
with our new algorithm. The line is more clearly identified, and subtle changes 
in direction are visible. The new algorithm is also 150x faster than the 
derivative-based method for this 128x128x64 array size (430s versus 3s). The 
scaling with larger array size is also much better. When tested on a 256x256x96 
array with 83,890 nonzero elements, the difference is 1,065s versus 4.7s, a 
speedup 
of 260x. This speedup makes incorporating dislocation identification into 
existing phase retrieval algorithms feasible. It could be used to perform the 
transformation to the dislocation basis, which tends to be sparse 
\cite{Tripathi2016}. Sparsity using the dislocation basis could be used to 
circumvent traditional constraints in BCDI, ptychography, and other methods that 
rely on phase retrieval. One may even be able to use subpixel resolution of 
angles to further refine dislocation core regions by using more sophisticated 
min-max algorithms such as that in \cite{MMSMW2017}.

We developed a new algorithm that is based on minimum differentials in a local 
neighborhood for identifying dislocation cores in BCDI images. Using 
experimentally determined images, we demonstrated that this algorithm is both 
more accurate and computationally efficient. Using the new algorithm, we identified additional 
geometric features of the dislocation line. The 
computational speedup opens the possibility of incorporating the dislocation 
line into the phase retrieval algorithm. For example, the image is sparse in the 
dislocation basis, and this feature could be exploited to develop new and improved 
algorithms for phase retrieval and new experimental techniques in which 
traditional constraints are relaxed. We expect this algorithm will find 
immediate use in identifying dislocations in reconstructed experimental images,
and we provide the algorithm in the supplemental material.

This work, including use of the Advanced Photon Source, was supported by the 
U.S.\ Department of Energy, Office of Science, Offices of Basic Energy 
Sciences (BES) and Advanced Scientific Computing Research (ASCR), under 
Contract No.\ DE-AC02-06CH11357. A.U.\ was supported by the BES Materials 
Sciences and Engineering Division; M.M.\ and S.M.W.\ were supported by the ASCR 
applied mathematics and SciDAC activities.  

\bibliography{../refs/smw-bigrefs}

\end{document}